# Flux dependent 1.5 MeV self-ion beam induced sputtering from Gold nanostructured thin films

J. Ghatak[1], B. Sundaravel[2], K. G. M. Nair[2], P. V. Satyam[1*]

[1]Institute of Physics, Sachivalaya Marg, Bhubaneswar, India, 751005

[2]Material Science Division, Indira Gandhi Center for Atomic Research, Kalpakkam, India, 603 102

**Abstact**

We discuss four important aspects of 1.5 MeV $Au^{2+}$ ion – induced flux dependent sputtering from gold nanostrcutures (of an average size ≈ 7.6 nm and height ≈ 6.9 nm) that are deposited on silicon substrates: (a) Au sputtering yield at the ion flux of $6.3\times10^{12}$ ions $cm^{-2}$ $s^{-1}$ is found to be ≈ 312 atoms/ion which is about five times the sputtering yield reported earlier under identical irradiation conditions at a lower beam flux of ≈$10^9$ ions $cm^{-2}$ $s^{-1}$ (at the fluence of $6\times10^{13}$ ions $cm^{-2}$) (b) the sputtered yield increases with increasing flux at lower fluence and reduces at higher fluence ($1.0\times10^{15}$ ions $cm^{-2}$) for nanostructured thin films while the sputtering yield increases with increasing flux and fluence for thick films (27.5 nm Au deposited on Si) (c) Size distribution of sputtered particles has been found to vary with the incident beam flux showing a bimodal distribution at higher flux and (d) the decay exponent (δ) obtained from the size distributions of sputtered particles showed an inverse power law dependence ranging from 1.5 to 2.5 as a function of incident beam flux. The exponent values have been compared with existing theoretical models to understand the underlying mechanism. The role of wafer temperature associated with the beam flux has been invoked for a qualitative understanding of the sputtering results in both the nanostructured thin films and thick films.

* Corresponding author: satyam@iopb.res.in, Tel: +91-674-2301058, Fax: +91-674-2300142

# 1. Introduction

Nanostructures and nanoparticles are grown by various physical methods such as molecular beam epitaxy, pulse laser deposition, sputter coating, etc. The sputter coating method involves the use of energetic particles (atoms or ions) bombarding the target of interest. During the sputtering process, the sputtered atoms or clusters generated due to the incident ion beam impingement condense on the surface of the specimen to be coated. Such processes are the basis of many thin film growth technologies (like DC magnetron or RF sputtering) [1]. Energetic ion beams have also been utilized in synthesizing and modifying nanostructures [2]. The ion irradiation being an athermal process, properties of nanomaterials could be tailored, which are otherwise difficult by conventional methods [3].

Among the many processes that occur during ion – solid interaction, sputtering plays a vital role as understanding the underlying mechanism of sputtering leads to a basic understanding of ion – solid interaction processes. It was established by Sigmund that the sputtering yield $Y$ (the average number of atoms released from a solid surface per incident particle), is proportional to the energy deposited by the projectile into the target through elastic collision processes (*i.e.*, nuclear energy loss of the projectile) [4]. But many experimental results obtained in the later years [5 – 8] indicate a clear deviation of the sputtering yield from the Sigmund's theory, such as, heavy ion bombardment on high – Z elemental targets. This has been attributed to the nonlinear sputtering processes. Many theoretical attempts have also been reported to understand the non-linearity effects in the sputtering process. One of the earlier attempts has come from Jonhson and Evatt [9] with a quantitative thermal spike model and then Sigmund and co-workers [10] with a modified thermal spike model. A qualitative hydro-dynamical model with an anisotropic velocity distribution [11, 12] and fluid dynamical analysis [13] were also proposed to explain the nonlinear sputtering. MD (molecular dynamics) simulations [14 – 18] have also been used to study the sputtering phenomena.

MD simulation study of sputtering of self-ion induced Au and Cu targets at keV energies clearly revealed the non-linearity in sputtering [15].

The sputtering studies show a monotonically decreasing yield distribution, which closely follows inverse power law decay

$$Y(n) = n^{-\delta} \qquad (1)$$

Where $n$ is the number of atoms present in the sputtered cluster. Shock wave model predicts the value of the exponent $\delta$ to be equal to 5/3 or 7/3 [19]. In an experiment involving 400 keV Au in Au film, Rehn et al. [20] obtained the value of $\delta$ to be around 2 for $n \geq 500$ and found to be consistent with the shock wave model [19]. This value of $\delta = 2$ is consistent with the mechanism that large clusters are produced when shock waves, generated by subsurface displacement cascades, ablate the surface (shock wave model). Recently, MD simulations have been carried out to study the effect of 100 keV Au bombardment on Au nanocluster of size 8 nm [17]. The results of the above MD simulation show a distribution of emitted clusters ($n \leq 100$). Smaller clusters (up to $n \sim 10$) show an inverse power law with $\delta=2.33$. This was explained in terms of the thermodynamic equilibrium description [21], which predicted the value of $\delta$ lie between 2 and 7/3. For larger clusters the value of $\delta$ is found to be higher than 7/3. These studies show the availability of various mechanisms for understanding the sputtering process.

Most of the reported experimental studies have been carried out on thick targets (i.e. not on nanostructured thin films) and in low keV energy regime. Recently, sputtering from the nanostructured Au films deposited on Si substrate has been investigated experimentally [22-26]. But no effort was made to study the power-law dependence of $Y(n)$. In one of our previous studies [25], the power-law dependence of the emitted larger particles ($n \geq 1000$) from Au nanostructures on silicon substrates due to MeV Au-ion bombardment at low ion flux condition ($1.3 \times 10^{11}$ ions cm$^{-2}$ s$^{-1}$) was reported. In this study, the decay exponent was found to be $\approx 1.0$ [25].

In this paper, we show that for the case of nanostructured thin films, the sputtering yield (Y) increases with the increasing of incident flux at lower fluence ($6\times10^{13}$ ions cm$^{-2}$) and reduces with incident flux at higher fluence ($1\times10^{15}$ ions cm$^{-2}$). These results are compared with the sputtering from thick films. The role of the morphology at the surface and interfaces under different incident flux conditions on the sputtering yield and the size distribution of the sputtered particles would be discussed. At high flux conditions, the transient wafer temperature would be higher and hence would play a role in the sputtering process.

## 2. Experimental

Au films of thicknesses 2.0 nm and 27.5 nm were deposited by resistive heating method in high vacuum conditions ($\approx 4 \times 10^{-6}$ mbar) and at room temperature on a $\approx$2 nm thick native oxide covered Si (111) substrates. Deposition rates for all samples were 0.01 and 0.1 nm/s for 2 nm and 27.5 nm thick Au films, respectively. Irradiations were carried out with 1.5 MeV Au$^{2+}$ ions at room temperature with incident ion beam flux values of $3.2\times10^{10}$, $6.3\times10^{11}$ and $6.3\times10^{12}$ ions cm$^{-2}$ s$^{-1}$ (hereafter these conditions will be referred as low flux (LF), medium flux (MF) and high flux (HF) respectively). The fluences on the samples were varied from $6\times10^{13}$ to $1\times10^{15}$ ions cm$^{-2}$. The substrates were oriented 5º off normal to the incident beam to suppress the channeling effect. The irradiations at a flux more than $1.3\times10^{11}$ ions cm$^{-2}$ s$^{-1}$ were carried out with 1.7 MV accelerator facility at Indira Gandhi Center for Atomic Research, Kalpakkam. The irradiation with beam flux less than $1.3\times10^{11}$ ions cm$^{-2}$ s$^{-1}$ and Rutherford Backscattering Spectrometry (RBS) have been performed using 3 MV accelerator facilities using a surface barrier detector of resolution 35 keV. All the irradiations were carried out using a raster scanner to have uniformity of irradiation. During the irradiation, the sputtered particles were collected on carbon coated copper grids (catcher grid) kept $\approx$1.0 cm above the sample, with a geometry as shown in Fig. 1(a). Care has been taken to have identical geometry for the collection of sputtered particles. RBS measurements with 2 and 3.0 MeV $He^{2+}$ ions were used to determine the variations in the effective film thicknesses of Au films before and after irradiation. Transmission electron microscopy (TEM) measurements were performed (using JEOL JEM-2010 operating at 200 KV) for the substrates before and after irradiations and on the catcher grids. Planar and cross-sectional TEM (XTEM) samples have been prepared using mechanical polishing followed by ion milling with 3.5 keV Ar ions.

## 3. Results and discussions

The irradiation experiments have been performed on Au nanostructured thin films of effective thickness ≈ 2.0 nm and 27.5 nm thick gold films deposited on Si(111) substrate. The effective thickness has been determined with SIMNRA simulation package [27] using the bulk Au density. In the RBS simulation (using SIMNRA), parameters like detector energy resolution and energy calibration values have been determined with the standard samples using bulk Si and thick Au film. For fitting the backscattered peak from Au, bulk density has been used and then the effective thickness was determined in the units of atoms $cm^{-2}$ (termed as areal density). With these fixed parameters, the RBS spectra of nanostructured thin films have been fitted using bulk Au density to obtain the area under the gold peak, which was used to determine the effective thickness. The sputtering yield was then obtained by dividing the areal density with the fluence. It should be noted that the surface and interface roughness would not affect the value of overall area under the curve and hence do not affect the value of the areal density. Prior to irradiation, the substrates were analyzed using planar and cross-sectional TEM. Figures 1 (b) and (c) represents bright field planar and cross-sectional TEM (XTEM) images of the pristine sample. Both the Figs. 1(b) and (c) show isolated Au islands that have been grown on the Si substrate (such thin films are termed as nano-structured thin films). The Figs. 1(d) and (e) represents the histogram of lateral size (from several frames like Fig. 1(b)) and height distribution (from several frames like Fig. 1(c)) of the Au nanostructures present on the Si surface. A Gaussian fit of the respective histogram gives an average Au nanostructure lateral size to be ≈ 7.6 ± 1.5 nm and the average height to be 6.9 ± 0.8 nm. Surface coverage for these nanostructured samples is found to be ≈ 40 %. The Fig. 1(f) depicts bright field XTEM micrographs of 27.5 nm thick gold film deposited on silicon. The thicknesses of Au films measured from both the RBS and TEM are in good agreement. It is evident from the cross-sectional micrographs that, a ≈2.0 nm thick native oxide was present at the interface of gold films and substrate. We present the detailed results on various aspects of sputtering from nanostructures in the following.

**(a) Sputtering yield measurements**

Figure 2(a) shows the RBS spectra obtained from the nanostructured Au films on Si targets before and after irradiation with 1.5MeV Au$^{2+}$ at fluence of 6×10$^{13}$ ions cm$^{-2}$ using the different flux values of 3.2×10$^{10}$, 6.3×10$^{11}$ and 6.3×10$^{12}$ ions cm$^{-2}$ s$^{-1}$ (LF, MF and HF conditions, respectively). From the RBS measurements, the reduction in the film thickness found to be 32%, 42% and 48% in LF, MF and HF conditions, respectively in comparison with pristine sample. The RBS measurements clearly indicate that with the increase of incident beam flux, the sputtering (reduction in amount of Au in substrate) also increases. From the RBS data of irradiated nanostructured Au films corresponding to a fluence of 6×10$^{13}$ ions cm$^{-2}$, the sputtering yield found to be as high as 125 atoms per ion for the HF condition, while the yield values were 107 and 95 atoms per ion corresponding to the MF and LF conditions. However the above sputtering yields were underestimated as the coverage of the Au islands was only 40% in the pristine film. By taking the coverage into account, the yields were found to be 237, 267 and 312 for LF, MF and HF condition, respectively. Figure 2(b) shows the RBS spectra obtained from nanostructured targets before and after irradiation with 1.5 MeV Au$^{2+}$ at a higher fluence (1×10$^{15}$ ions cm$^{-2}$) as a function of incident flux. At this high fluence, the thickness reductions in the film thicknesses were found to be 94%, 64% and 45% at LF, MF and HF conditions, respectively. It is interesting note that the total thickness reduction values (in % value) at fluence of 6×10$^{13}$ ions cm$^{-2}$ (for LF condition is 32% while for HF condition is 48%) and at 1×10$^{15}$ ions cm$^{-2}$ (for LF condition is 94% while for HF condition is 45%). Determining the values of the sputtering yield at the high fluence conditions (for nanostructured films) would be highly inaccurate to use 40% coverage area. The coverage area of gold at intermediated fluences was not measured and hence absolute values of Y would not be presented. The observed effects of beam flux on sputtering yield and variations at the high fluence will be discussed in later parts by considering the mass transport phenomena from Au films into the Si substrate. Figure 2(c) shows the RBS spectra obtained from 27.5 nm thick continuous Au film on Si before and after irradiation with 1.5 MeV Au$^{2+}$ at fluence of 1×10$^{14}$ ions cm$^{-2}$ for MF and HF conditions. For these systems, the reductions in the Au film thicknesses were found to be 12% and 22% at MF and HF conditions, respectively. The reduction in the thickness increases with increase of beam flux for thick target, which is similar to that, has been seen Fig. 2(a). The yield for the thick film has also been calculated in similar manner as observed for nanostructure film systems and the corresponding sputtering yield values at MF and H F conditions are 144 and 340, respectively.

The Table I show some of the sputtering yield data available in the literature (theoretical, simulated and experimental) and from the present work for 1.0 – 2.5 MeV self – ion induced sputtering from gold films. Among the existing theories on nonlinear sputtering, shock wave model based on hydrodynamical analysis [11, 12] fits well with many of the experimental results. But the theory gives over-estimated sputtering yield at energies more than 1.0 MeV. Most of the experimental work has been carried out at low flux and fluence to avoid cascade-overlapping and prominent non-linear effects. It should be noted that our present work is different from other in two aspects: (a) sputtering from nanostructures and (b) sputtering as a function of flux (low to high flux conditions). From the table I, it is evident that a sputtering yield of about 70 was observed from previous experimental observations [6 – 8]. This value is about five times less than the sputtering yield observed for high incident beam flux conditions both for nanostructured and continuous gold films for 1.5 MeV incident ions. It is to be noted that the sputtering yield for both type of targets came out to be comparable at HF condition, whereas sputtering yield at lower flux is less in thick films compared to nanostructured films.

Recent studies by the group of Baranova et al showed large sputtered yields for nano-dispersed targets with $Au_5$ cluster ions with energy 200 keV/atom (low flux, low fluence and nuclear energy loss dominant regime) [23]. Whereas, under similar irradiation conditions (i.e. with cluster ions), the sputtering yield from bulk gold targets was found to be less compared to nano-dispersed systems [7]. Interestingly, the sputtering yield calculated from SRIM [28] is 27 (as mentioned in Table I) which is low compared to the sputtering yield ($\approx$ 48) calculated from Sigmund's linear theory [4] for 1.5 MeV Au$\rightarrow$ Au. SRIM results also show lower yield for thinner films in contradiction to the recent experimental results [23, 24]. Though it is true that sputtering phenomena is significantly dominated in nuclear energy loss regime, recent studies show the contribution of electronic energy loss to the sputtering yield is prominent [26, 29]. In the present work, the ratio of electronic to the nuclear energy loss is 0.26 for 1.5 MeV $Au^{2+}$ in Au target *i.e.* electronic energy loss contributes 21% to the total energy loss and hence should not be neglected in the sputtering calculation.

The prominent variation in sputtering for nanostructured films and thick films at high fluence ($1\times10^{15}$ ions cm$^{-2}$) and HF condition is due to the variation in ion beam induced interface mixing in Au/Si systems. To understand the surface and interface morphology in irradiated systems XTEM measurements have been carried out. Figure 3(a) depicts a XTEM micrograph of the nanostructured target after irradiation at fluence $1\times10^{14}$ ions cm$^{-2}$ under LF condition. From this figure, it is evident that, no interface mixing has been taken place. We can also infer that no interface mixing takes place for fluence less than $1\times10^{14}$ ions cm$^{-2}$ under low flux and normal incident conditions ($5^0$ impact angle). At HF condition, an unusual mass transport from the nanostructured film was found to be present. More details of the mass transport under high flux condition have been discussed elsewhere [30, 31]. Fig. 3(b) shows a XTEM bright field micrograph of nanostructured target after irradiation at fluence of $1\times10^{14}$ ions cm$^{-2}$ irradiated at HF condition. Inset of Fig. 3(b) shows a high resolution (HR) lattice image from a region shown as circle in Fig. 3(b). From this figure, it is clear that Au atoms from the nanostructures on Si surface have been transported to the interface and reacted to form gold silicde. The HR lattice image shows a spacing of $0.305 \pm 0.005$ nm. As the Si substrate has already amorphized and no gold d-spacing matches with this value, we concluded that a metastable gold silicide ($Au_5Si_2$) has been formed [30,]. Figures 3 (c), (d) and (e) show the XTEM image of irradiated nanostructured system after irradiation at fluence of $1\times10^{15}$ ions cm$^{-2}$ under LF, MF and HF conditions, respectively. At this fluence, a very large amount of material has been sputtered out at LF condition. This is consistent with the 94% reduction in RBS spectra (as shown in Fig. 2(b)). At the fluence of $1\times10^{15}$ ions cm$^{-2}$ and under HF condition, surface craters and interface mixing has been observed (Fig. 3(e)). Unlike the absence of large amount of Au at surface in case of LF condition, there appears to be more amount of Au available for sputtering at the interface. Presence of craters may also play a role in reduction of sputtering as well. At MF condition, there is more material present on the surface when compared to both LF and HF conditions. Under similar conditions, irradiation effects from thick film have also been studied and depicted in Fig. 3(f). The Fig. 3(f) shows XTEM image of thick film after irradiated at a fluence of $1\times10^{14}$ ions cm$^{-2}$ under HF condition. It is to be noted that at low flux, no interface mixing has been observed for thick films [32], while at high flux (HF condition) mass transport across the interface is evident from Fig. 3(f). Even though mass transport across the interface has been observed for the thick film under HF condition, there is enough Au material is present on the surface and hence more sputtering at higher fluence has been observed. But this is not the case for nanostructured target. It is expected that the embedded Au has a lower sputtering yield due to dilution and the small surface area. At the high fluence conditions, as whole amount of Au at the surface has been sputtered (also for the LF conditions), very little sputtering would be seen. In other words, the lack of surface Au leads to a slowing of the sputtering in case high flux and high fluence irradiation conditions.

It is clear from the above experimental observations that the incident ion beam flux plays an important role in sputtering process. In the following, the role of high flux condition in terms of rise in the transient wafer temperature has been discussed. As given in by Nakata's formalism [33], the wafer temperatures have been calculated for the fluxes during irradiations (detail calculation has been reported elsewhere [30]). At the highest flux used in the present study (i.e., $6.3 \times 10^{12}$ ions $cm^{-2}$ $s^{-1}$) and for the fluence $6 \times 10^{13}$ ions $cm^{-2}$, the temperature calculated using the prescription of Nakata would be 1125 K (for an irradiation time of 9 s). Following the above calculations, for the fluence $6 \times 10^{13}$ ions $cm^{-2}$ at a flux of $6.3 \times 10^{11}$ ions $cm^{-2}$ $s^{-1}$, the temperature would be 650 K (irradiation time: 90 s) and the wafer temperature would be 400 K (irradiation time: 460 s) for $1.3 \times 10^{11}$ ions $cm^{-2}$ $s^{-1}$. This means that at same fluence, the temperature of the wafer during irradiation is higher for the higher beam flux. As the wafer temperature increases, the heat of sublimation (ΔH) of Au decreases and hence the binding energy also decreases. It has already been established that sputtering yield is inversely proportional to the binding energy [4, 34]. Hence, a rise in the wafer temperature results in enhancement of sputtering cross-section. Sigmund and Szymonsky (1984) reviewed the temperature dependent scenario up to that period and theoretically predicted that high temperature regime (thermal spike) yields little variation in sputtering yields [35]. But the experiments and simulations studies by Behrisch et al., showed a drastic Ag sputtering yield enhancement with increasing target temperature during irradiation [34]. Increase of sputtering (as long as enough material is present) on the substrate in both thick and nanostructured Au films at fluence and at HF condition has been attributed to the wafer temperature.

**(b) Sputtered particle size distribution:**

The sputtered particle size distribution as a function of beam flux is discussed below. The sputtered particles have been collected (geometry is shown in Fig. 1(a)) on carbon coated grid. Figures 4(a), (b) and (c) shows the TEM micrograph of sputtered Au particles collected on catcher grid at LF, MF and HF conditions respectively, at a fluence of $6\times10^{13}$ ions cm$^{-2}$. Figures 4(d), (e) and (f) are the corresponding size distribution of sputtered particles whose TEM data has been shown in Figs. 4(a), (b) and (c), respectively. To give quantitative information about the sputtered particles, we have fitted the particle size distribution with the log-normal distribution which is given by:

$$f(x) = \frac{1}{\sqrt{2\pi}\, xw} \exp\left(-\frac{(\ln(x/x_c))^2}{2w^2}\right) \qquad (2)$$

Where $x_c$ and $w$ are the most probable size and the width of the size distribution, respectively. This is because the distribution of nanoparticles often found to be the log-normal distribution [37]. The average particle size was found to be 7.7 ± 0.3 and 9.5 ± 0.2 nm for samples in LF and MF conditions (Figs. 4(d) and 4(e)), respectively. Interestingly, a bimodal distribution has been found in HF irradiation condition (as shown in Fig. 4(f)) with average particle sizes of 3.4 ± 0.3 nm and 10.0 ± 0.2 nm. It is also to be noted that the width (*w*) of the distribution is minimum under HF irradiation condition. At higher fluence ($1\times10^{15}$ ions cm$^{-2}$) under HF conditions also a bi-modal distribution of sputtered particles for the same system has been observed [38]. The origin of the bimodal distribution has been explained as follows: At a flux and fluence values of $6.3\times10^{12}$ ions $cm^{-2} s^{-1}$ and $6\times10^{13}$ ions $cm^{-2}$, Au islands reacted with Si and formed gold silicide at the interface [30]. That means, at higher beam flux, the Au islands on the Si undergo ion beam mixing and form a silicide phase whereas in lower beam flux at this fluence, no ion beam mixing has taken place across the interfaces. At the initial conditions of irradiation, the availability of gold material is more compared to a later stage (for the case of beam incidence on nanostructures). The size of the sputtered particle is appeared to be directly proportional to the amount available in the nanostructures. This results in sputtering of bigger clusters at initial stages and smaller clusters at later stage, giving rise to a bimodal distribution (as shown in Fig. 4(f)).

**(c) Sputtering yield decay exponent ($\delta$):**

The value of decay exponent ($\delta$) would help in understanding the underlying mechanism of sputtering process [20]. The sputtered particles size analysis has been carried to determine the decay exponent [25]. Assuming a spherical nature for the sputtered particles, the distribution obtained using the particle size diameter has been converted into the hemi-spherical volume distribution as given by Rehn et al. [25]. Figure 5 depicts the values of *Y(n)* versus hemispherical volume of the sputtered particles (which is directly proportional to the value of *n*. In the Fig. 5, the legends S1, S2 and S3 corresponding to the irradiated samples at the fluence of $6\times10^{13}$ ions $cm^{-2}$ in HF, MF and LF conditions respectively. To avoid overlap in plotting, S1, S2 and S3 were multiplied by factors 8.0, 1.0 and 0.25, respectively. The data points marked as S1, S2 and S3 were fitted with a straight line to obtain the $\delta$ values. For the hemispherical volume more than 100 $nm^3$, from a straight line fit (as shown in Fig. 5), $\delta$ values found to be $2.0 \pm 0.1$ and $1.5 \pm 0.1$ for S2 and S3 respectively. Because of bimodal distribution (Fig. 4 (f)) for S1, we got two $\delta$ values, $1.6 \pm 0.2$ for the particle sizes $\leq 5$ nm and $2.5 \pm 0.2$ for particle sizes $\geq 6$ nm. The average sputtered particle size is more for sample S1 and hence faster decay has been taken place. The δ value for sample S3 is nearly same as that was reported earlier [25]. The shock wave model predicts the δ value to be 2 or 2.33 [19] and may be comparable with values from S1 and S2. If shock wave mechanism assumed to be the underlying mechanism, then the same δ values for the irradiation at varying beam fluxes would be expected. Thus from present data, it is clear that the sputtering mechanism is not the same at all incident ion beam fluxes. Though, the shock wave model works for many experimental results [20, 39] under ion irradiation conditions at low flux on thick continuous films, the model is not an appropriate for studying the sputtering from nanostructured targets.

A proper reasoning for the different exponent values is difficult in the present experimental conditions due to the presence of complicated surface and interface morphology, mass transport and alloy-formation across the interfaces, sublimation of nanostructures, and multi – ion impacts [30]. The MD simulations of Kissel and Urbassek [17] showed the sputtering from spherical Au clusters (radius $R$ = 4 nm) due to 100 keV Au atom bombardment. For the smaller cluster sizes $n \leq 10$, the data indeed follow a polynomial decay with $\delta \cong 2.3$. While comparing this result with the MD simulation study of 100 keV Au bombardment of Au(111) surface (>12 000 atoms), the decay exponent comes out to be $\delta \cong 2.8$ [16]. From the above studies it can be inferred that large clusters are emitted in the case of spherical island system with a higher probability than for planar solid surfaces. It also clearly indicates that the mechanism of sputtering from nanostructured target is different from that from a solid surface (continuous layer). For a nanostructured target, sputtering was attributed to the thermodynamic equilibrium description [17] where cluster production from a volume energized to reach the critical point of the gas–liquid phase transition is known to lead to a cluster-mass distribution $\propto n^{-7/3}$ [21]. In our present experiments, only decay exponent of samples irradiated at highest flux, $6.3 \times 10^{12}$ ions cm$^{-2}$ s$^{-1}$, only is close to the decay exponent of thermodynamic equilibrium description.

## 4. Conclusions

In conclusion, the present results conclusively show that the sputtering yield is affected by the incident ion beam flux. With the ion flux of $6.3 \times 10^{12}$ ions cm$^{-2}$ s$^{-1}$, the sputtering yield of Au from a nanostructured target due to 1.5 MeV Au ions has been found to be as high as 312 atoms/ions which is comparable with the values for thick continuous film. Our experimental data on the size of sputtered particles at a given fluence with different beam flux values clearly shows a strong influence of the ion beam flux on the sputtered particle size distribution. Decay exponent of sputtering particles is also found to be varying from $\approx 1.5$ to 2.5 for the flux values from $3.2 \times 10^{10}$ to $6.3 \times 10^{12}$ ions cm$^{-2}$ s$^{-1}$ suggesting that the mechanisms operative at different flux values are different. The models that describe these effects should take the beam flux effects. The higher sputtering at higher beam flux has been attributed to beam induced transient temperature raise.

## Acknowledgements

We acknowledge the cooperation of S. Amirthapandian during high flux ion irradiation experiments. We would also like to thank all the staffs of ion beam laboratories at IOP and IGCAR.

**Figure captions:**

**Figure 1:** (Color online) (a) depicts the schematic diagram of the collection arrangement of sputtered particles on to carbon coated grids.(b) and (c) are the planar and XTEM images of pristine 2.0 nm Au deposited on Si, respectively. (d) and (e) correspond to the size and height distributions of the Au islands present on the Si surface determined using planar and XTEM measurements. Both the histograms are fitted with Gaussian distribution just to get the values of average size and height of the Au islands present on the substrate. (f) corresponds to the XTEM images of pristine 27.5 nm Au deposited on Si.

**Figure 2:** (Color online) (a) and (b) correspond to the RBS spectra for nanostructured targets (2.0 nm Au on Si) before and after irradiation at a fluence $6\times10^{13}$ and $1\times10^{15}$ ions cm$^{-2}$ at various flux values. (c) the RBS spectra for thick continuous targets (27.5 nm Au/Si system) before and after irradiation at fluence of $1\times10^{14}$ ions cm$^{-2}$ at two different fluxes. The RBS spectra were collected using 3 MeV He$^+$ for (a) and (b) while a 2 MeV He$^+$ ion beam was used for (c). LF, MF and HF in figures denote the ion beam flux $3.2\times10^{10}$, $6.3\times10^{11}$ and $6.3\times10^{12}$ ions cm$^{-2}$ s$^{-1}$ respectively.

**Figure 3:** (a) and (b) correspond to the bright field XTEM micrograph for nanostructured target (2 nm Au on Si) after irradiation with flux $3.2\times10^{10}$ and $6.3\times10^{12}$ ions cm$^{-2}$ s$^{-1}$ respectively at fluence of $1\times10^{14}$ ions cm$^{-2}$. The inset of (b) is the high resolution lattice image has taken from the circular part of (b). (c) - (e) correspond to the XTEM images of irradiated nanostructured targets with flux of $3.2\times10^{10}$, $6.2\times10^{11}$ and $6.3\times10^{12}$ ions cm$^{-2}$ s$^{-1}$ at a fluence of $1\times10^{15}$ ions cm$^{-2}$. (f) is the XTEM image of the thick continuous targets (27.5 nm Au film on Si) after irradiation at a fluence of $1\times10^{14}$ ions cm$^{-2}$ and flux of $6.3\times10^{12}$ ions cm$^{-2}$ s$^{-1}$.

**Figure 4:** (Color online) (a) – (c) Bright-field TEM micrographs of sputtered Au nanoparticles have been collected on catcher grids during irradiation of nanostructured target (2 nm Au on Si) with 1.5 MeV Au$^{2+}$ at a fluence 6×10$^{13}$ ions cm$^{-2}$. (a)–(c) correspond to the ion beam flux 3.2×10$^{10}$, 6.3×10$^{11}$ and 6.3×10$^{12}$ ions cm$^{-2}$ s$^{-1}$ with corresponding size distributions shown in (d)–(f) respectively. The histograms in (d), (e) and (f) have been fitted with a log–normal distribution function (as mentioned in text). The most probable size, $X_c$, and width, $W$, of the size distribution have indicated in the figure.

**Figure 5:** (Color online) The log-log plot of hemisphere volume distribution. S1, S2 and S3 correspond to at a fluence 6×10$^{13}$ ions cm$^{-2}$ with flux 6.3×10$^{12}$, 6.3×10$^{11}$ and 3.2×10$^{10}$ ions cm$^{-2}$ s$^{-1}$. To avoid overlap, S1, S2 and S3 were multiplied by 8, 1 and 0.25. Straight solid lines are linear fit to curves S1, S2 and S3 to get the decay constants (δ) for each beam flux.

**Table I:** Some of theoretical and experimental sputtering yields for Au→ Au from previous works and present data. SRIM values are also tabulated to compare the values. The references denotes the theoretical (Th), SRIM simulation (SS) and experimental (Ex) observations. LF, MF, HF denotes the irradiation conditions of high flux, medium flux and low flux corresponding to values of 3.2×10$^{10}$ ions cm$^2$ s$^{-1}$, 6.3×10$^{11}$ ions cm$^2$ s$^{-1}$, and 6.3×10$^{12}$ ions cm$^2$ s$^{-1}$).

**Table I**

| Energy (keV) | Film thickness (nm) | Sputtering Yield | Reference |
|---|---|---|---|
| 1500 | - | 48 | 4 [Th] |
| 1500 | - | 63 | 11 [Th] |
| 1500 | - | 106 | 12 [TH] |
| 1500 | 2 | 12 | 28 [SS] |
| 1500 | 200 | 27 | 28 [SS] |
| 1000 | 300-600 | 80 | 6 [Ex] |
| 2500 | 300-600 | 62 | 6 [Ex] |
| 1400 | 1000 | 65 | 8 [Ex] |
| 1500 | 2.0 | 312 [HF] / 267 [MF] / 237 [LF] | This work [Ex] |
| 1500 | 27.5 | 340 [HF] / 144 [MF] | This work [Ex] |

**Figure1:**

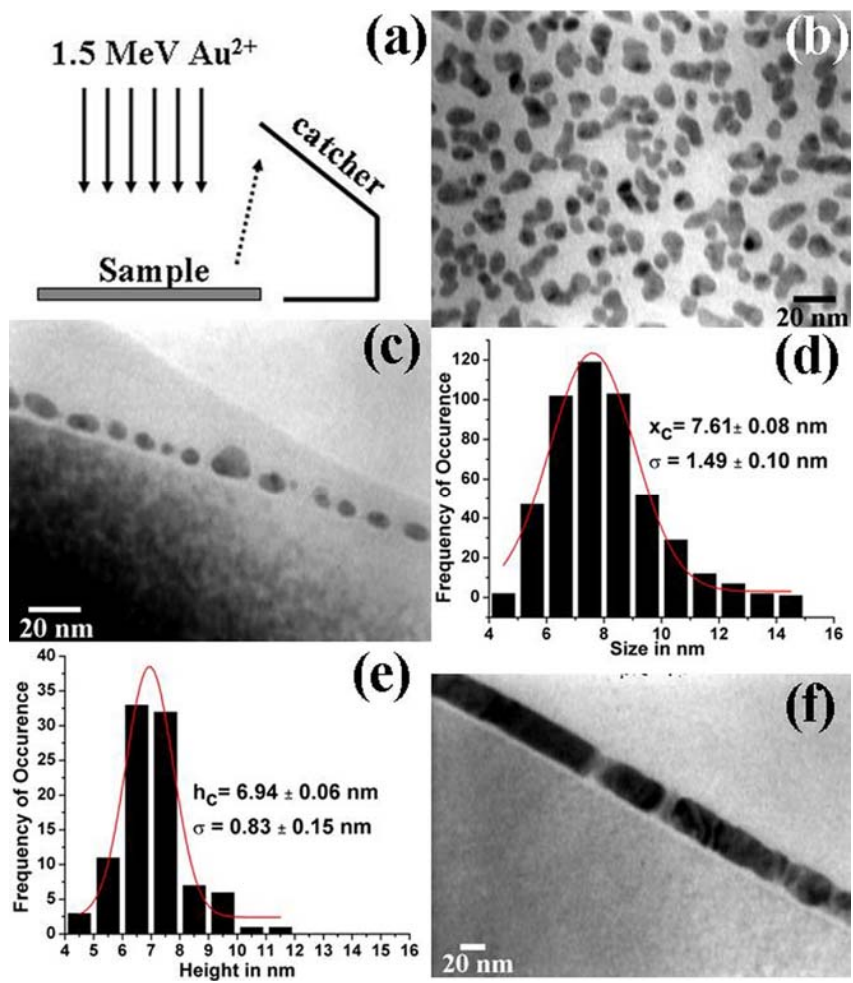

**Figure2:**

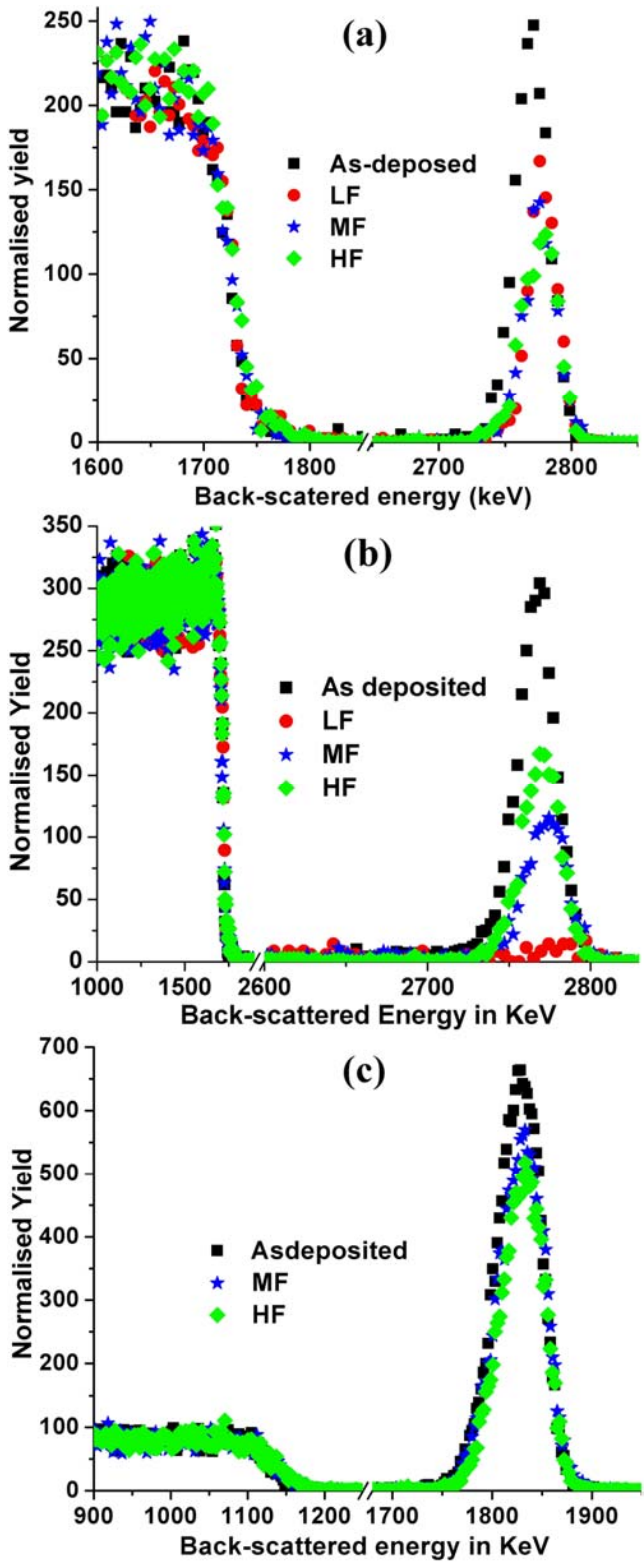

**Figure3:**

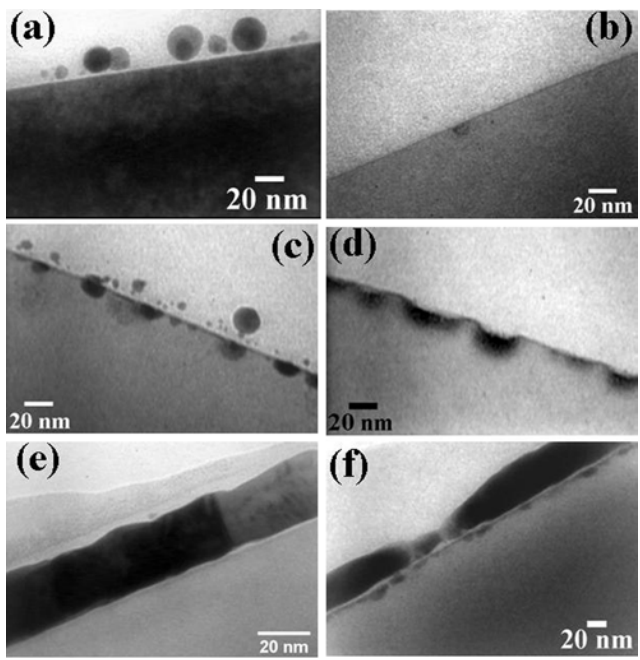

**Figure4:**

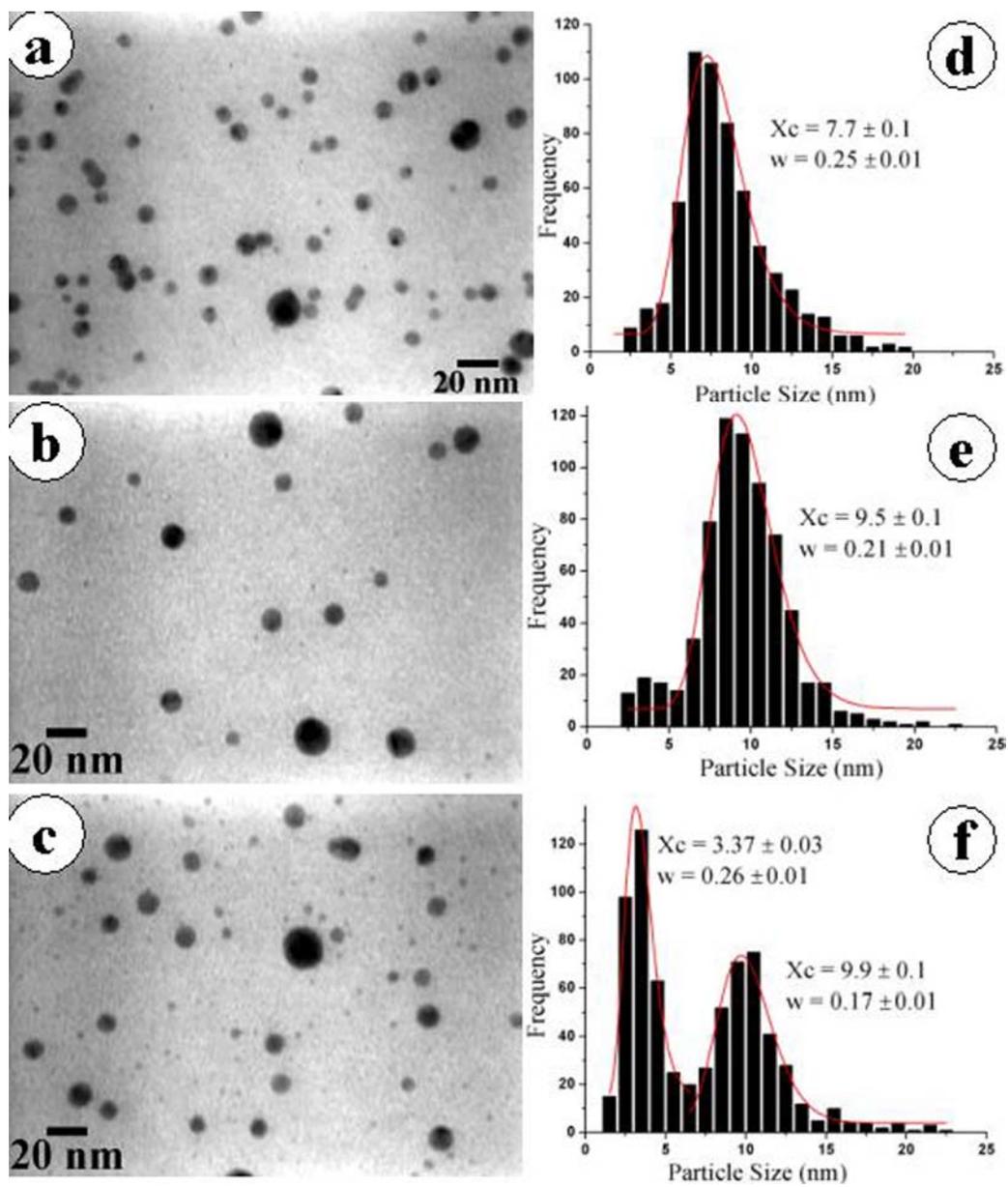

**Figure5:**

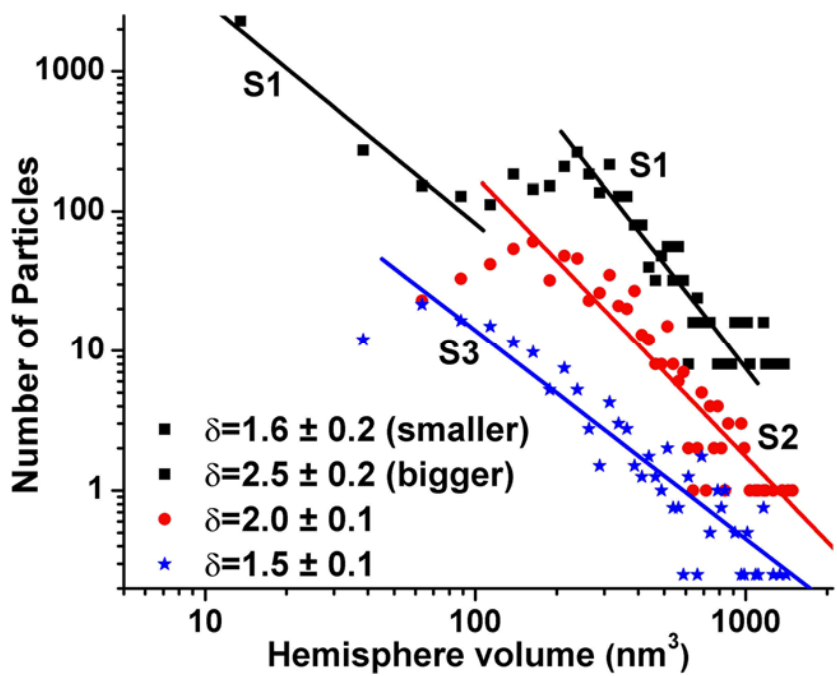